\documentclass[aps,prd,twocolumn,superscriptaddress,amsfont,graphicx,nofootinbib,preprintnumbers]{revtex4}%
\usepackage{color,graphicx,epsfig}
\usepackage{ifpdf}
\usepackage{amsmath}
\usepackage{bm}
\usepackage{color}
\usepackage[english]{babel}
\usepackage{graphicx}%
\usepackage{amsfonts}%
\usepackage{amssymb}
\usepackage{braket}
\usepackage{hyperref}
\usepackage{enumerate}

\definecolor{nicered}{rgb}{0.7,0.1,0.1}
\definecolor{nicegreen}{rgb}{0.1,0.5,0.1}
\hypersetup{colorlinks,citecolor= nicegreen,linkcolor= nicered}

\newcommand{\beq}{\begin{equation}}
\newcommand{\eeq}{\end{equation}}
\newcommand{\bea}{\begin{eqnarray}}
\newcommand{\eea}{\end{eqnarray}}

\definecolor{Red}{rgb}{1.,0.,0.}

\def\OMIT#1{}

\begin{document}

\def\Fermilab{Fermilab, P.O.Box 500, Batavia, IL 60510, USA}
\def\Durham{Institute for Particle Physics Phenomenology, Department of Physics, \\ University of Durham, Durham, DH1 3LE, UK}
\def\Buffalo{Department of Physics, University at Buffalo \\ The State University of New York, Buffalo 14260 USA}

\preprint{
\noindent  IPPP/17/25, 
FERMILAB-PUB-17-085-T}
\flushbottom

\title{Driving Miss Data: Going up a gear to NNLO}

\author{John M. Campbell}     
\email[Electronic address:]{johnmc@fnal.gov}
\affiliation{\Fermilab}

\author{R. Keith Ellis}     
\email[Electronic address:]{keith.ellis@durham.ac.uk}
\affiliation{\Durham}

\author{Ciaran Williams}     
\email[Electronic address:]{ciaranwi@buffalo.edu}
\affiliation{\Buffalo}

\date{\today}
\begin{abstract}
In this paper we present a calculation of the $\gamma+j$ process at next-to-next-to-leading order (NNLO) in QCD
and compare the resulting predictions to 8 TeV CMS data. We find good agreement with the shape of the photon $p_T$ spectrum,
particularly after the inclusion of additional electroweak corrections, but there is a tension between the overall normalization of the theoretical
prediction and the measurement. We use our results to compute the ratio of $Z(\to \ell^+\ell^-)+j$ to $\gamma+j$ events as a function of the vector boson transverse
momentum at NNLO, a quantity that is used to normalize $Z(\rightarrow\nu\overline{\nu}) +j$ backgrounds in searches for dark matter and supersymmetry.
Our NNLO calculation significantly reduces the theoretical uncertainty on this ratio, thus boosting its power for future searches of new physics. 
\end{abstract}

\maketitle

\section{Introduction} \label{sec:intro}

One of the primary aims of the LHC's physics mission is to search for Beyond the Standard Model (BSM)  physics. A key
motivation for BSM physics arises from the cosmological observations of Dark Matter (DM).  Thus far, multiple
observations have inferred the existence of DM through its gravitational interactions with baryonic matter (see
ref.~\cite{Bergstrom:2012fi} for a recent review);  however
to date no observation of non-gravitational interactions of DM has been conclusively established.   The search for
non-gravitational interactions of DM is hence an ongoing and exciting area of active research. 

At the LHC the putative DM particle, or any similarly weakly-interacting BSM state, will not be directly observed by the LHC detectors.
Instead the particle may be pair-produced in association with jets, that are observed in copious amounts at the LHC.  If the DM particle couples
to the SM through a heavy mediator then the typical transverse energy of the DM pair will be large, with the jets accounting for the corresponding
recoil in the transverse plane.  This would allow the presence of the DM to be inferred from an excess of events with large missing transverse
energy (MET).  As a result the MET+jets channel is one of the most exciting and rich channels in which to search for BSM effects
(for a recent overview see ref.~\cite{Askew:2014kqa}).  

Unfortunately, the Standard Model (SM) itself also provides a substantial source of events with large MET.  The largest source of such events
is through the production of a $Z$-boson in association with jets, with the subsequent decay $Z\rightarrow \nu\overline{\nu}$.
Since the invisible decay forbids the reconstruction of the invariant mass of the parent $Z$-boson,
this background cannot be easily suppressed by an explicit mass-window cut.
This presents a significant challenge for MET+jet searches. Thankfully the visible decays of the $Z$ provide a window through which to study this 
irreducible background~\cite{Bern:2011pa,Ask:2011xf}. Decays of the $Z$ boson to light charged leptons, $Z\rightarrow e^+e^-$ and $Z\rightarrow \mu^+\mu^-$, are clean experimental signatures with 
excellent resolution. By studying the impact of artificially not taking into account the visible leptons, the effect of
the transition to MET-based observables can be easily quantified.  However, a secondary issue arises when using the charged leptons as a tool to measure the neutrino background. Since the branching ratio for $Z\rightarrow \ell^+\ell^-$ 
is significantly smaller than for $Z\rightarrow \nu\overline{\nu}$ there are considerably less $Z\rightarrow{\ell}^+\ell^- +$jets events than MET+jets ones. 
At high vector boson transverse momentum ($p_T^V$), exactly the region of most interest, the low statistics of the  $Z\rightarrow{\ell}^+\ell^-$ mode
limits its utility for estimating the $Z\rightarrow \nu\overline{\nu}$ background. 

In the region of high $p_T^V$ one must therefore find an alternate strategy for calibrating the MET+jets background. One possibility is 
to make use of the sample of $\gamma$+jet events. The photon and $Z$ boson are similar enough that a comparison of their production mechanisms is useful and, since one
does not have to pay the price of a branching ratio for the photon, there is a factor of $\mathcal{O}(\alpha_{ew}^{-1}) \sim  100$ more events at high $p_T$.
One can therefore measure the ratio of ${\ell}^+\ell^- $+jets and $\gamma+$jets events
at low $p_T$ and extrapolate into the high $p_T^Z$ region. A good agreement between theory and data for this ratio is
crucial; only once it has been demonstrated at lower values of $p_T^V$ can the method be applied with confidence in the region of limited data at
higher values of $p_T^V$.

Theoretical predictions for the $Z+j$ and $\gamma+j$ processes have been available at NLO for a long time~\cite{Giele:1993dj,Catani:2002ny}.
From these calculations the theoretical uncertainty associated with a truncation of the perturbative expansion at this
order may be estimated from the sensitivity of the
predictions to the choice of factorization, renormalization and (in the case of $\gamma+j$) fragmentation scales.  These are typically in the
range of $10$--$20$\%, which has been sufficient for testing the SM in these channels in the past.  However, as the LHC accumulates more data
of this nature~\cite{Khachatryan:2010fm,Aad:2013zba,Chatrchyan:2013mwa,Aad:2016xcr}, the experimental uncertainties are approaching the level
of a few percent and will only decrease further.  In order to achieve a similar level of theoretical precision it is necessary to include
additional perturbative corrections.  For the case of $Z+$jet production, NNLO QCD corrections have been extensively studied by
now~\cite{Ridder:2015dxa,Boughezal:2015ded,Ridder:2016nkl,Boughezal:2016isb}.  At this level of accuracy it
is also necessary to include the effect of NLO electroweak corrections, which are also known for this process~\cite{Kuhn:2004em,Kuhn:2005az}.
For $\gamma+$jet production, the closely-related direct photon process has recently been computed at NNLO in QCD~\cite{Campbell:2016lzl} and
the NLO EW corrections are known as well~\cite{Kuhn:2005gv}.

In this paper we will provide NNLO predictions for $\gamma+$jet production, thus bringing the theoretical prediction
to the same level as for the $Z+$jet process.
To do so we will make use of the direct photon calculation of Ref.~\cite{Campbell:2016lzl}, that has already been implemented in the
Monte Carlo code MCFM, and explicitly demand the presence of a jet.  With this calculation in hand we will
be able to address the main aim of this paper, which is predicting the ${\ell}^+\ell^- +$jet$/\gamma+$jet ratio with an accounting of NNLO
QCD and leading EW effects.  To do so we will also make use of the MCFM implementation of the NNLO corrections to
$Z+$jet production~\cite{Boughezal:2015ded}.

   
\section{Calculation} \label{sec:Calc}
 
 \subsection{IR regularization}
 
NNLO calculations require regularization of infrared singularities that are present in phase spaces with 
different numbers of final state partons.  In our
calculations we use the $N$-jettiness slicing approach that was outlined in refs.~\cite{Boughezal:2015dva,Gaunt:2015pea}, based on
earlier similar applications to top-quark decay at NNLO~\cite{Gao:2012ja}. This method follows a divide-and-conquer approach
to regulating the singularities in the calculation. A cut on the $N$-jettiness variable $\tau_N$~\cite{Stewart:2010tn}
is introduced, where $N$ is the number of jets in the Born phase space. For the case at hand $N=1$. 
Therefore we introduce the following variable 
\begin{eqnarray}
\label{eq:tau1def}
\tau_1 = \sum_{k=1}^{M}\min_{i=a,b,1}\left\{\frac{ 2 q_i \cdot p_k}{Q_i}\right\} \;.
\end{eqnarray}
Where $\{p_k\}$ defines the momenta of the parton-level configuration, and $\{q_i\}$ represents the set of momenta that is obtained after
application of a jet-clustering algorithm.  The scale $Q_i$ is a measure of the jet or beam hardness, which we take as $Q_i = 2E_i$.  
The labels $a$ and $b$ refer to the two beam partons.
Note that if
$\tau_1 =0$ then the clustered momenta map directly onto the Born phase space (i.e. a one-jet configuration). Non-zero values of $\tau_1$
therefore correspond to configurations with a greater number of partons than
the Born phase space.  We introduce a cut choice $\tau^{\rm{cut}}_1$ such that when
$\tau_1 > \tau^{\rm{cut}}_1$ the components of the calculation 
contain at most single-unresolved infrared singularities. It therefore corresponds to a NLO calculation with
an additional parton, albeit one which must be integrated with an extremely loose jet requirement. The double-unresolved
singularities reside in the region $\tau_1 < \tau^{\rm{cut}}_1$, where the application of
SCET~\cite{Bauer:2000ew,Bauer:2000yr,Bauer:2001ct,Bauer:2001yt,Bauer:2002nz} allows us to write the cross section as follows, 
\begin{eqnarray}
\sigma(\tau_1  < \tau^{\rm{cut}}_1) =
 \int \mathcal{H} \otimes  \mathcal{B}  \otimes  \mathcal{B}   \otimes  \mathcal{S}  \otimes  \mathcal{J} + \mathcal{O}(\tau^{\rm{cut}}_1) \,.
\label{eq:belowcut}
\end{eqnarray}
That is, the cross section factorizes into a convolution of process-independent beam ($\mathcal{B}$) and jet ($\mathcal{J}$) functions, a soft
function $\mathcal{S}$ (which depends on the number of colored scatterers) and a (finite) process-specific hard function $\mathcal{H}$. Expansions
accurate to $\mathcal{O}(\alpha^2_s)$, that are relevant for our calculation, can be found in refs.~\cite{Gaunt:2014xga,Gaunt:2014cfa},
~\cite{Becher:2006qw,Becher:2010pd} and~\cite{Boughezal:2015eha} for the beam, jet and soft functions respectively. The hard functions for the
processes we consider in this paper are written in terms of the two-loop virtual matrix elements that have been calculated in
ref.~\cite{Anastasiou:2002zn} and refs.~\cite{Gehrmann:2011ab,Gehrmann:2013vga} for the $\gamma+$jet and $Z+$jet cases respectively. 
Their implementation has been discussed in ref.~\cite{Boughezal:2015ded} for $Z+j$ production and in ref.~\cite{Campbell:2016lzl} for
direct photon production, which shares the same hard function as the photon+jet case we consider here.
A key consideration within the $N$-jettiness slicing approach is the choice of $\tau^{\rm{cut}}_1$ used for the calculation.  As indicated in
Eq.~(\ref{eq:belowcut}), the below-cut factorization theorem receives power corrections that vanish in the limit $\tau^{\rm{cut}}_1\rightarrow 0$,
but they can have a sizable impact on the cross section for non-zero values. Therefore it is crucial that $\tau^{\rm{cut}}_1$ be taken as small as
possible, to minimize the impact of these corrections.\footnote{For recent work on reducing the dependence on power corrections, see
refs.~\cite{Moult:2016fqy,Boughezal:2016zws}.} A general discussion of the process-specific parts of the direct photon and $Z+$jet calculations
in MCFM was presented in refs~\cite{Campbell:2016lzl,Boughezal:2015ded}. For brevity we will not reproduce that discussion here, but refer the
interested reader to the original works for further details. Instead, in this paper we will focus on the validation of both calculations for
the specific phase space selection criteria employed by the CMS analysis that we will follow.

\subsection{Parameter choices}
The usual MCFM EW parameter choice is the $G_\mu$ scheme, in which the values of $M_W$,
$M_Z$ and $G_\mu$ (the Fermi constant) are taken as inputs. In this scheme the electromagnetic
coupling is then defined, at leading order, as 
\begin{eqnarray}
\alpha_{G_\mu} = \frac{G_\mu M_W^2 \sqrt{2}}{\pi} \left(1-\frac{M_W^2}{M_Z^2}\right)
\end{eqnarray}
A disadvantage of this scheme for our calculation, which involves real photons in the final state,
is that this choice of $\alpha$ ($\sim \!\! 1/132$) is rather large compared to the fine-structure constant
($\alpha_{(0)}\sim 1/137$) that is more appropriate for on-shell photons.
Converting to the $\alpha_{(0)}$ scheme, in which $\alpha_{(0)}$ is taken as an input rather than $G_\mu$,
has its own disadvantages though: it induces larger higher-order electroweak corrections and, via
renormalization, introduces a dependence on light-quark masses. We therefore follow
ref.~\cite{Alioli:2016fum} and work in a modified $G_\mu$ scheme in which only the LO couplings
are expressed in terms of $\alpha_{G_\mu}$, with higher-order corrections evaluated at $\alpha_{(0)}$.
An additional advantage of this choice is that the dependence on $\alpha_{G_\mu}$ partially cancels
in the $Z+j$/$\gamma+j$ ratio.\footnote{There is still a residual dependence on
$\alpha_{G_\mu}$ from the $Z\rightarrow \ell^+\ell^-$ decay.}
Our calculations are thus performed using the following parameters:
\begin{eqnarray}
&& M_Z = 91.1876\,\mbox{GeV} \,, \qquad \Gamma_Z = 2.4952\,\mbox{GeV} \,, \nonumber \\
&& M_W = 80.385 \,\mbox{GeV} \,, \qquad \sin^2\theta_w = 0.222897 \,, \nonumber \\ 
&& \alpha(M_Z) = 1/132.232 \,.  
\end{eqnarray}
We will choose both renormalization ($\mu_R$) and factorization ($\mu_F$) scales equal to $H_T$,
which is defined event-by-event to be the scalar sum of the transverse momenta of all particles
present.  When studying the theoretical uncertainty associated with this choice of scale we consider
a six-point variation corresponding to,
\begin{equation}
\mu_R = r H_T \,, \qquad \mu_F = f H_T \,,
\label{eq:scalevariation}
\end{equation}
with $r, f \in (\frac{1}{2}, 1, 2)$ and $rf \ne 1$.  We use the NNLO CT14 set of parton distribution
functions~\cite{Dulat:2015mca}.  Studies
of the associated PDF uncertainty are performed using the additional 56 eigenvector sets provided
through LHAPDF6~\cite{Buckley:2014ana} and are quoted at the 68\% confidence level.

\subsection{Event selection}
\label{sec:eventselection}

Our phase space selection criteria are based on those used in a recent CMS analysis of 8 TeV data~\cite{Khachatryan:2015ira}.
For the photon plus jets sample we require that the photon satisfies the following cuts 
\begin{eqnarray}
p_T^{\gamma} > 100 \;\;{\rm{GeV}} \,, \quad |\eta_{\gamma}| < 1.4 \,.
\end{eqnarray}
Both experimentally and theoretically photons require isolation from hadronic activity. On the experimental side this reduces unwanted backgrounds from 
pion decays and photons that arise from fragmentation processes. Theoretically the calculation is simplified if smooth cone isolation~\cite{Frixione:1997ks} is employed. 
In that case one requires that the photon satisfies 
\begin{eqnarray}
\sum  p_T^{\rm{had}}(R) < \epsilon_{\gamma} p_T^{\gamma} \left(\frac{1-\cos{R}}{1-\cos{R_0}}\right)^n \quad \forall R < R_0 \,.
\label{eq:smoothiso}
\end{eqnarray}
This requirement constrains the sum of the hadronic energy inside a cone of radius $R$, for all separations $R$ that are smaller than a
chosen cone size, $R_0$.  
Cones are defined in terms of the $R$ variable,
\beq
R  =  \sqrt{\Delta \eta^2 + \Delta \phi^2} \,,
\eeq
where $\eta$ and $\phi$ are the pseudorapidity and azimuthal angle of the particle, respectively.
Note that arbitrarily soft radiation will always pass the condition, but collinear $(R \rightarrow 0)$ radiation is
forbidden.  This removes the collinear splittings associated with fragmentation functions, at the cost of no longer reproducing the form
of isolation applied in experimental analyses.  In this paper we set $\epsilon_{\gamma} =0.025$, $R_0=0.4$ and $n=2$ in Eq.~(\ref{eq:smoothiso}).
This matches the parameters employed in a similar analysis by the BlackHat collaboration~\cite{Bern:2011pa}.\footnote
{We note that these parameters are slightly different to those used in previous MCFM studies of photonic processes at
NNLO~\cite{Campbell:2016lzl,Campbell:2016yrh}. We  have compared with the alternative choice $\epsilon_{\gamma} =0.1$
and found that the cross section only changes by around 1\%.}
At NLO we can explicitly quantify the difference between
following this procedure and performing a calculation that includes the effects of fragmentation.  We shall
see later that this difference is small, around a percent in the photon $p_T$ spectrum. 
 
In addition to the photon requirements described above, we require the presence of at least one jet in the
event. Jets are defined using the anti-$k_T$~\cite{Cacciari:2008gp} algorithm  with $R=0.5$ and satisfy,
\begin{eqnarray}
p_T^{\gamma} > 30 \;\;{\rm{GeV}} \,, \quad |\eta_{j}| < 2.4 \,.
\end{eqnarray}
Additionally we require that photons and jets are separated by $R_{\gamma j} > 0.5$. 

For the $Z+j$ sample we require that the charged leptons are in the following fiducial volume,
\begin{eqnarray}
p_T^{\ell} > 20 \;\;{\rm{GeV}} \,, \quad |\eta_{\ell}| < 2.4 \,.
\end{eqnarray}
We require that the lepton pair resides in an invariant mass window close to the $Z$ mass, $71 < m_{\ell\ell} < 111$ GeV, and that the
leptons are isolated from jets, $R_{\ell j} > 0.5$. We also require that $p_T^Z > 100$ GeV and $|y_Z| < 1.4$ to mimic the photon selection
as closely as possible.

\subsection{$\tau_1^{\rm{cut}}$ dependence }
 
\begin{center}
\begin{figure}
\includegraphics[width=8cm]{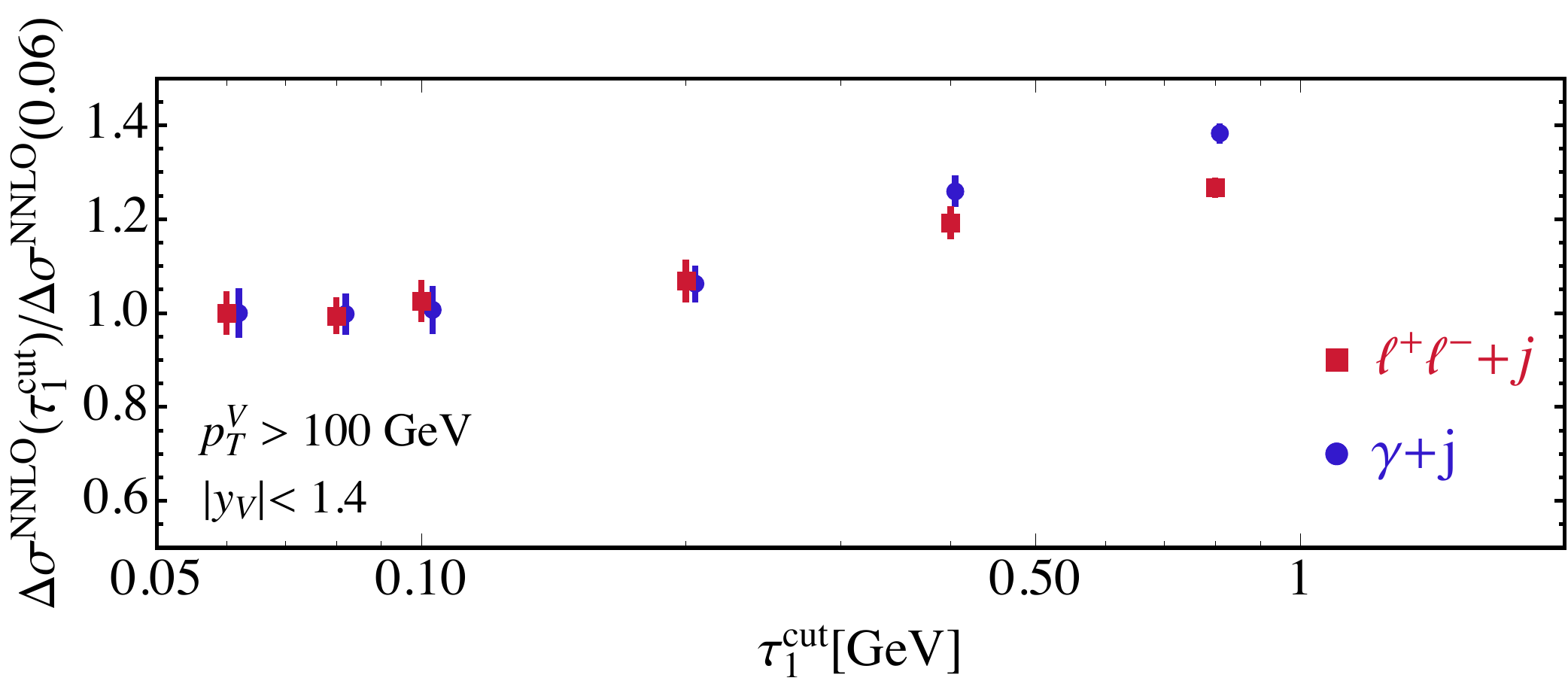}
\caption{The dependence of the NNLO coefficient on the parameter $\tau^{{\rm{cut}}}_1$ for the processes considered in this paper.
The cuts of the CMS analysis~\cite{Khachatryan:2015ira} have been applied.
To aid visibility, the values of $\tau^{{\rm{cut}}}_1$ for the $\gamma+j$ calculation have been offset slightly. }
\label{fig:taudep}
\end{figure}
\end{center}

Before providing NNLO predictions for $\gamma +j$ (and $Z+j$) production
we first validate our calculation for the
phase space cuts described in the previous section. Since the $N$-jettiness slicing method is sensitive to
power corrections it is crucial to validate the calculation for a new phase space selection. At NNLO the
cross section can be written as 
\begin{eqnarray}
\sigma_{NNLO} &=& \sigma_{NLO} +  \Delta \sigma_{NNLO},
\end{eqnarray}
where $\sigma_{NLO}$ is the NLO cross section and $\Delta \sigma_{NNLO}$ represents the correction
that arises at NNLO.  In MCFM,
$\sigma_{NLO}$ is calculated using a traditional Catani-Seymour dipole subtraction method~\cite{Catani:1996vz} and only $\Delta
\sigma_{NNLO}$ is computed using $N$-jettiness slicing. Therefore only $\Delta \sigma_{NNLO}$ has a
dependence on $\tau^{\rm{cut}}_1$, a sensitivity that is indicated in Fig.~\ref{fig:taudep}. This figure shows
the ratio $ \Delta \sigma_{NNLO}(\tau_1^{\rm{cut}})/ \Delta \sigma_{NNLO} (\tau_1^{\rm{cut}} = 0.06$ GeV),
for both of the processes considered in this paper.  Since the cuts have been chosen to emphasize the similarity
between the two processes we see that, as expected, the dependence on $\tau_1^{\rm{cut}}$ is also comparable.
Below $\tau_1^{\rm{cut}} = 0.1$ GeV the predictions are insensitive to the choice of
$\tau_1^{\rm{cut}}$ within Monte Carlo uncertainties which, in this region, are around
5\%.\footnote{We note that the MC uncertainties are all rescaled by the central value at
$\tau_1^{\rm{cut}} = 0.06$ GeV such that there is no reduction in uncertainties due to the fact that the plotted
quantity is a ratio.} We will see that $\Delta \sigma_{NNLO}/\sigma_{NNLO}$ is approximately 5--10\% for both
processes, so that the resulting uncertainty on $\sigma_{NNLO}$ due to power corrections and Monte Carlo statistics
is below 1\%.  This is perfectly acceptable for phenomenological purposes and, given the results in
Fig.~\ref{fig:taudep}, we choose $\tau_1^{\rm{cut}} = 0.08$ GeV to compute the remainder of the results in this paper. 

\subsection{Electroweak corrections}

Since datasets at the LHC now permit the study of $\gamma+$jet and $Z+$jet events in which the photon or $Z$-boson
carries a transverse momentum approaching 1~TeV, it is imperative to also account for the effect of electroweak
corrections in theoretical predictions for these processes.  Although these are generically expected to be rather
small, at such high transverse momenta they give rise to Sudakov-enhanced corrections of the order of $10$\% or
more.  These primarily arise from the contribution of loop diagrams in which a virtual $W$- or $Z$-boson is
exchanged, resulting in leading logarithms of the form $\log^2(M_V/p_T)$, whose effects on these processes
have been known for some time~\cite{Kuhn:2004em,Kuhn:2005az,Kuhn:2005gv}.  More recently these effects have also
been computed in the framework of SCET, which also allows an inclusion of terms corresponding to mixed
QCD-electroweak corrections~\cite{Becher:2013zua,Becher:2015yea}.

In this paper we shall make use of the results of Refs.~\cite{Becher:2013zua,Becher:2015yea} in order to
account for electroweak effects.  In these papers the effect of the electroweak corrections is captured
by expressing their effect on the cross section ($\sigma_{EW}$) as a fraction of the leading order result,
\begin{equation}
\Delta_{\rm{EW}} = \frac{\sigma_{EW}}{\sigma_{LO}} \,.
\end{equation}
$\Delta_{\rm{EW}}$ is then parametrized as a
function of the transverse momentum of the $Z$-boson or photon and the center-of-mass energy,
$\sqrt{s}$. This simple parametrization is expected to be robust against the application of mild
experimental cuts such as the ones used in this paper. We note that the authors of 
Refs.~\cite{Becher:2013zua,Becher:2015yea} used a value of $\alpha \sim 1/128$ which should be altered to
$1/137$ in our modified $G_\mu$ scheme. However, since this is a $7\%$ correction on what is itself
at most about a $10\%$ correction, this difference manifests itself in sub-percent effects.
We therefore take the results from Refs.~\cite{Becher:2013zua,Becher:2015yea} without modification and 
tolerate the discrepancy. For the $\gamma+$jet process we have explicitly checked that this approach agrees with the
one-loop NNLL results presented in Ref.~\cite{Kuhn:2005gv} (evaluated with $\alpha_{(0)}$) up to negligible
numerical differences.  We will treat the EW corrections as factorizing fully with respect to the QCD ones and simply
multiply our NNLO QCD predictions by $1+\Delta_{\rm{EW}}$.

\section{Differential predictions for $\gamma + j$}

\begin{center}
\begin{figure}
\includegraphics[width=8cm]{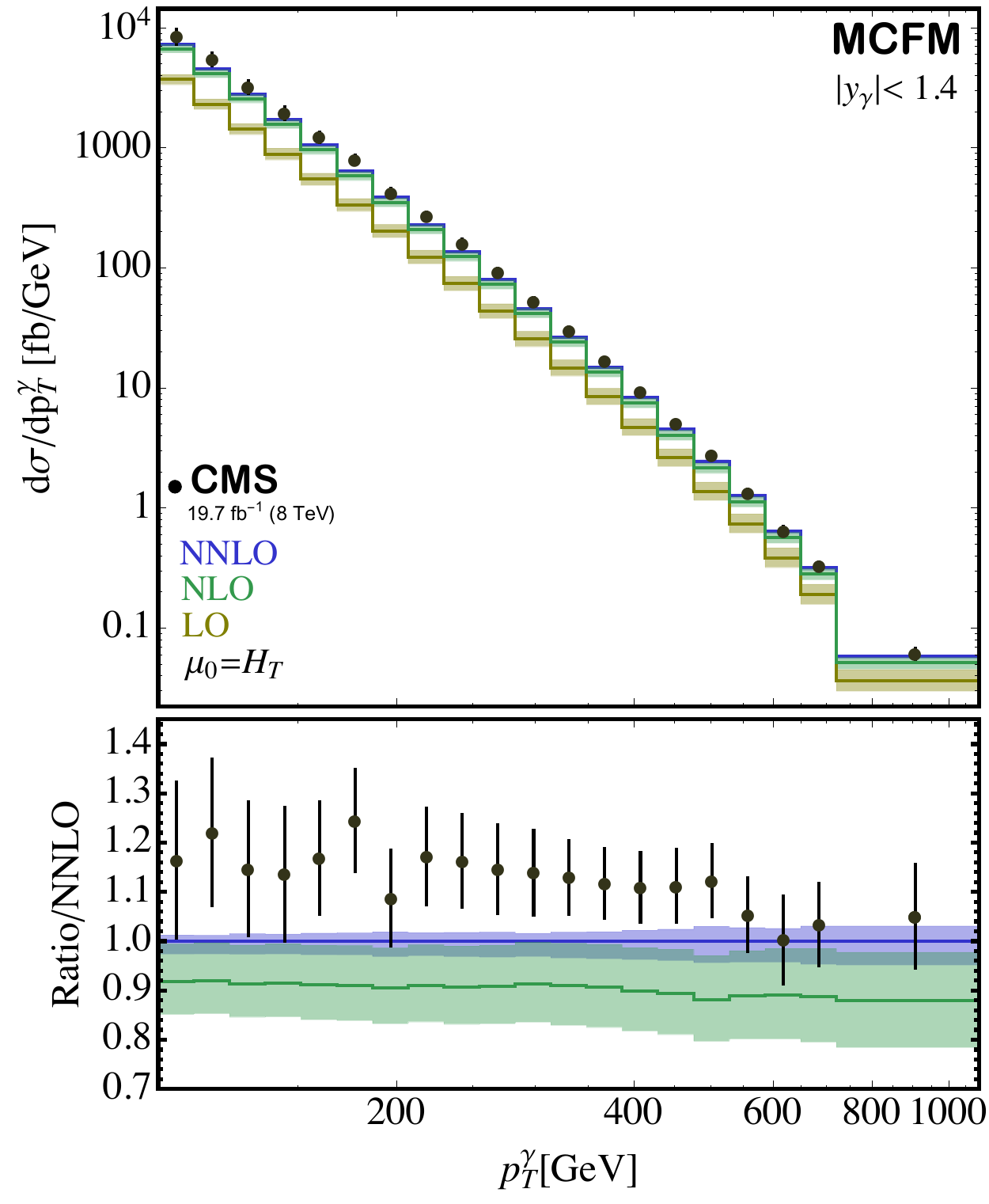}
\caption{The photon $p_T$ spectrum for $\gamma+j$ at the 8 TeV LHC, at various orders in
perturbation theory, compared to CMS data from ref.~\cite{Khachatryan:2015ira}. The lower
panel shows the ratio of the data and the NLO prediction to the NNLO one.
The bands indicate the scale uncertainty on the NLO and NNLO predictions.}
\label{fig:ptgam}
\end{figure}
\end{center}
 
Before arriving at the primary interest of this paper, an analysis of the $Z+j$/$\gamma+j$ ratio at NNLO, we first consider the
$\gamma+j$ process on its own. As discussed in the introduction, the $Z+j$ process has been extensively studied at NNLO,
including detailed phenomenological analyses~\cite{Ridder:2015dxa,Boughezal:2015ded,Ridder:2016nkl,Boughezal:2016isb}.
No such studies exist for the $\gamma+j$ process at this order and a careful analysis is a prerequisite to studying
the ratio in detail.  
Therefore in this section we compare the predictions of MCFM for $\gamma+j$ production to CMS data collected at 8 TeV. 
The fundamental quantity of interest is the photon transverse momentum spectrum, which we present in Fig.~\ref{fig:ptgam}.
The correction from NLO to NNLO is around 10\% and the NNLO prediction lies just at the very top of
the scale variation band obtained at NLO.  The NNLO/NLO $K$-factor is reasonably flat, with a slight increase
at higher  $p_T$. The scale variation at NNLO is significantly reduced compared to that obtained at
NLO, with a typical variation of $2$-$3$\% compared to $8$-$10\%$ at NLO. Although
the NNLO prediction lies closer to the CMS data than the NLO one, both predictions are consistently
lower than the experimental measurements.

\begin{center}
\begin{figure}
\includegraphics[width=8cm]{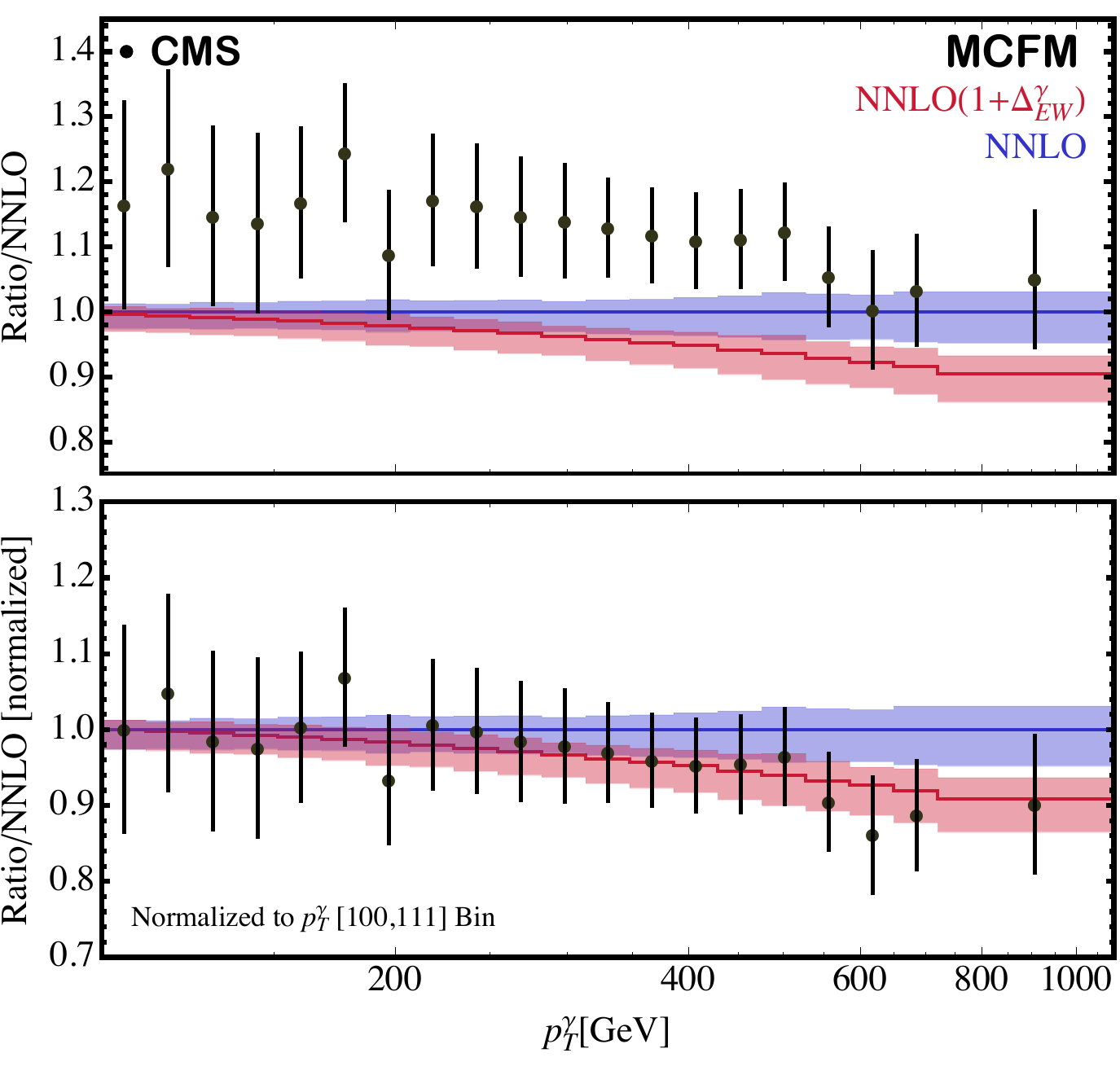}
\caption{The ratio of the CMS data from ref.~\cite{Khachatryan:2015ira} to the NNLO prediction (with and without
including EW effects) for the photon $p_T$ spectrum.  The lower panel normalizes this ratio to the value of the
ratio in the $[100,111]$ bin.}
\label{fig:ptgaNorm}
\end{figure}
\end{center}

We now include the effect of electroweak corrections as discussed above, by rescaling the complete NNLO result
by the change observed in the LO prediction when including one-loop electroweak effects.  We denote this combination
by the shorthand NNLO(1+$\Delta_{\rm{EW}}$).  Fig.~\ref{fig:ptgaNorm} shows the ratio of data and NNLO(1+$\Delta_{\rm{EW}}$)
to the pure NNLO prediction for the photon $p_T$ spectrum. The upper panel shows the raw ratio, while the lower panel
normalizes all predictions to their central value in the $p_T^{\gamma} \in [100,111]$ GeV bin, allowing us
to compare the shape of the predictions. We note that this procedure results in an overestimate of the
errors on the CMS data, since a normalized distribution should not be sensitive to the overall luminosity.
However, for the purposes of this comparison this overestimate can be tolerated. However, a full analysis of the
shape of the distribution measured by the LHC collaborations and a comparison to its theory counterparts is clearly very desirable.
The upper panel shows that, by including the EW corrections, the apparent agreement between theory and
data gets worse. However, the  lower panel shows that the shape of the data and theory predictions are
actually in very good agreement.

We have so far only considered the theoretical uncertainty originating from the choice of scale
and demonstrated that it is significantly reduced at NNLO, by a factor of two.  However there are other origins
of theoretical uncertainty, beyond scale variation, that affect our prediction.  We will now
consider three other sources of theoretical uncertainty: PDFs, choice
of $\alpha_{EM}$ and the form of the photon isolation.  These may primarily affect
the normalization of the theoretical prediction, or may induce changes in the shape
of the distributions.  For PDF uncertainties we will consider the 68\% confidence level
uncertainties provided by LHAPDF6~\cite{Buckley:2014ana} where, for efficiency, these
uncertainties are computed from the NLO prediction (using NNLO CT14 PDFs). We have checked that
the difference in PDF uncertainty obtained from LO and NLO calculations using this set is very
small, so that we are confident that this provides a reliable estimate of the
PDF uncertainty for our NNLO prediction. In addition we consider the change in the
overall normalization induced by excursions from our choice of $\alpha_{EM}$,
corresponding to the extreme choices of the $\alpha_{(0)}$ scheme or of choosing a higher-scale
value, $\alpha_{EM} = \alpha_{EM}(M_Z) = 1/127.9$.  Note that, since the running of $\alpha_{EM}(Q)$ is very slow for $Q \gtrsim M_Z$, this choice is practically equivalent
to a dynamic choice such as $\alpha_{EM}(p_T^\gamma)$.   In order to quantify the effect of the difference
between our isolation prescription and that of the experimental analysis, we repeat our
NLO calculation using the parton-level version of the experimental isolation procedure:
\begin{equation}
E_T^{\rm{had}} < 5\,\mbox{GeV} \quad \forall R < R_0 \,.
\end{equation}
Here, as in the smooth cone version, $R_0 = 0.4$ and our calculation employs the GdRG
fragmentation functions~\cite{GehrmannDeRidder:1997gf}.  Since the difference between the methods of
isolating the photon could be affected differently at NNLO we should provide a conservative estimate
of this effect. We therefore estimate the isolation uncertainty by taking the difference between
the two isolation procedures, multiplying by an additional factor of two, and allowing excursions from
our central result by this amount on either side.


\begin{center}
\begin{figure}
\includegraphics[width=8cm]{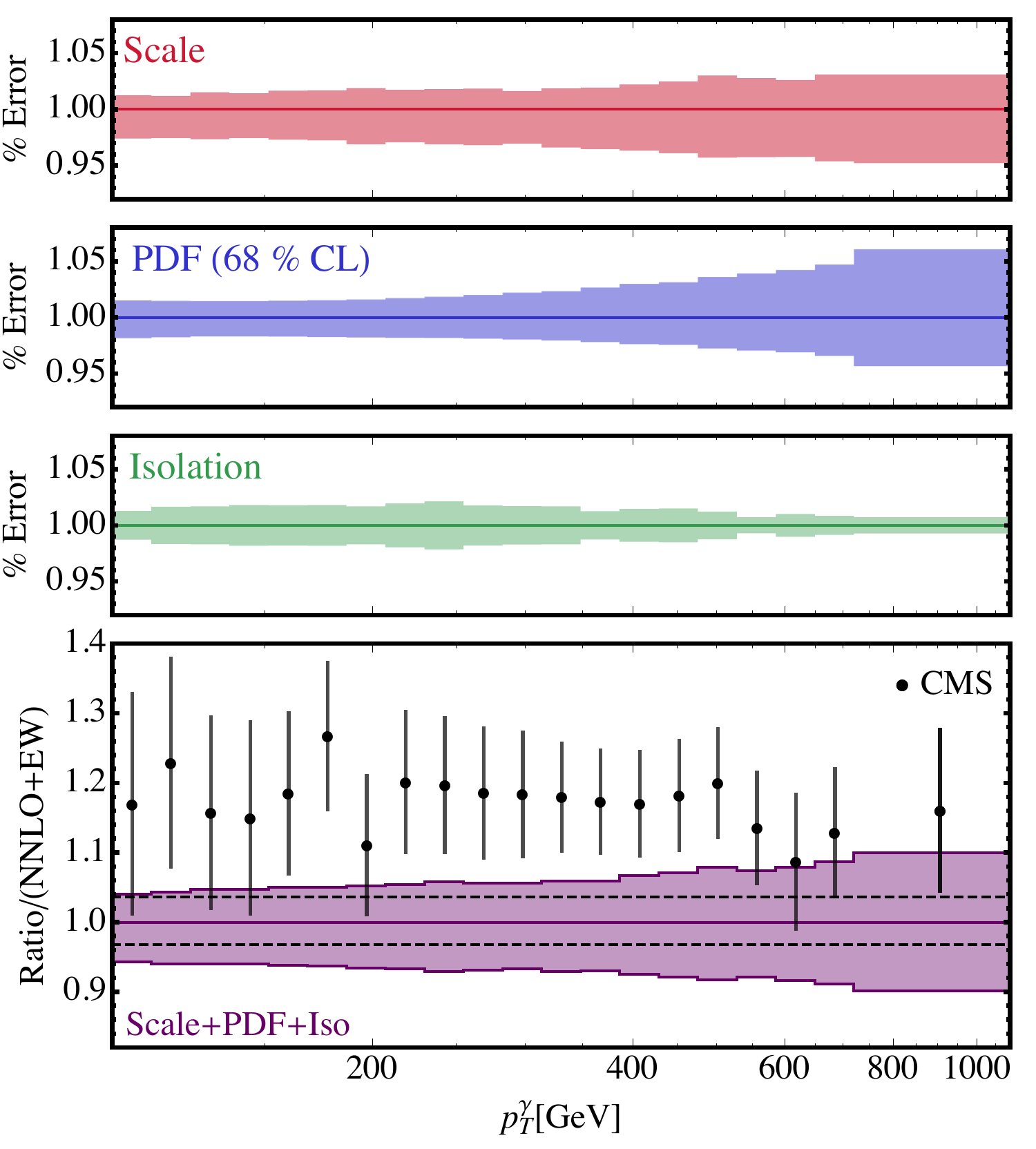}
\caption{A summary of the theoretical uncertainties discussed in this paper for the photon
transverse momentum spectrum.  In order from the top, uncertainties from: scales, PDFs, isolation
and in the total, as described in the main text. The total uncertainty
is obtained by combining linearly those from the sources above.  The uncertainty due to
the value of $\alpha_{EM}$ is indicated separately by the dashed line in the lower figure.}
\label{fig:pdfuncert_ptgam}
\end{figure}
\end{center}

Our results for the uncertainty in the theoretical prediction for the photon $p_T$ spectrum
are presented in Figure~\ref{fig:pdfuncert_ptgam}. The uncertainties are normalized to the
central value of the combined NNLO QCD + NLO EW prediction.  We observe that at NNLO
the scale variation and PDF uncertainty are roughly equal  and correspond to a few percent uncertainty.
The PDF uncertainty grows more rapidly as a function of photon transverse
momentum and is largest in the highest bins ($\sim 5\%$).
The uncertainty stemming from the isolation procedure is at the level of $2\%$ for lower values of
$p_T^\gamma$ but is significantly smaller in the tail.  This is in line with previous studies of
the difference between smooth cone isolation and the forms used in experimental
analyses~\cite{Catani:2013oma,Campbell:2016lzl}. 
The total uncertainty from scales, PDFs and isolation,  obtained by adding the individual uncertainties linearly,
ranges from around 4\% at low $p_T^\gamma$ to 9\% in the highest bins.  We separately indicate the 
normalization  uncertainty, due to the value of $\alpha_{EM}$, which is competitive with the other sources of
uncertainty at low $p_T^\gamma$.  Clearly the large PDF uncertainty can be reduced in the
future~\cite{dEnterria:2012kvo,Carminati:2012mm}, by taking advantage of calculations such as this one in tandem
with the even bigger $\gamma+j$ data sets being accumulated at the LHC.



The tension that remains between the data and our theoretical prediction, displayed in the lower panel of 
Figure~\ref{fig:pdfuncert_ptgam}, could have a number of sources.  Although we have endeavored to be thorough,
the accounting of theoretical uncertainty could yet be deficient.   
On the experimental side the normalization of the data could be changed by a host of factors, including
a reduction in the overall luminosity, a change in the photon efficiency, or an issue with background subtraction.\footnote{We note that the
CMS paper~\cite{Khachatryan:2015ira} indicates a flat 2.6\% luminosity uncertainty over the whole $p_T^\gamma$ range,
which is far below the level of disagreement indicated here.}

\begin{center}
\begin{figure}
\includegraphics[width=8cm]{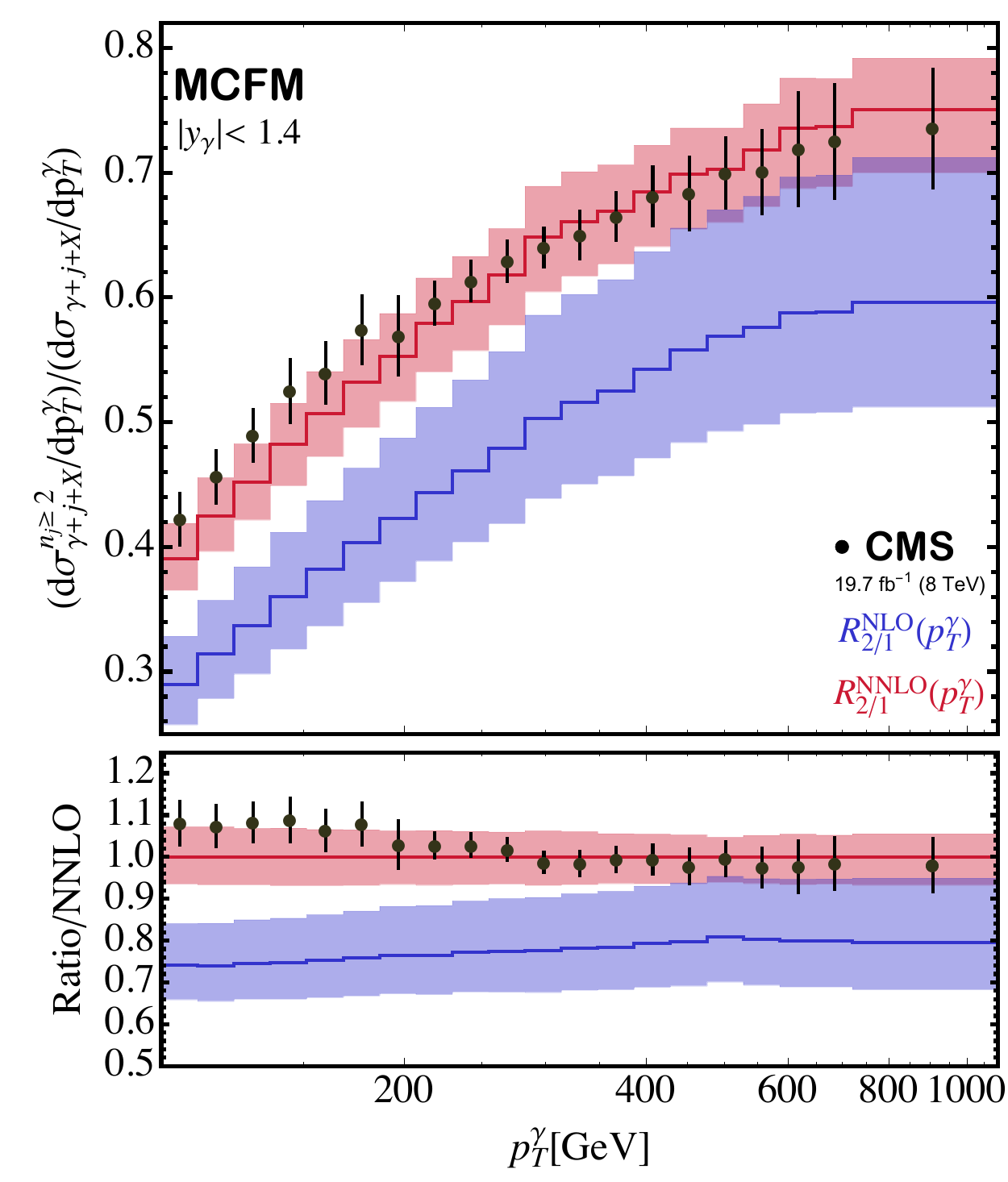}
\caption{The quantities $R_{2/1}^{\rm{NLO}}(p_T^\gamma)$ and $R_{2/1}^{\rm{NNLO}}(p_T^\gamma)$
compared to CMS data from ref.~\cite{Khachatryan:2015ira}.
The bands indicate the scale uncertainty on the theoretical predictions.}
\label{fig:njetratio}
\end{figure}
\end{center}

A further interesting observable to consider is the ratio of inclusive
$\gamma + 2 j$ to $\gamma +  j$ production as a function  of the photon transverse momentum.
Fixed-order calculations of this ratio can be broken down into contributions proportional
to the relevant powers of the strong coupling as follows,
\begin{equation}
 R_{2/1}(p_T^\gamma) = 
 \frac{\alpha_s^2 \sum_{k=0}^{n_2} \alpha_s^k \, d\sigma_{\gamma+2j}^{(k)}/dp_T^\gamma}
      {\alpha_s   \sum_{k=0}^{n_1} \alpha_s^k \, d\sigma_{\gamma+j}^{(k)}/dp_T^\gamma} \,.
\label{eq:R21defnew}
\end{equation}
In this expression we have made clear that contributions to the denominator start with one
power of $\alpha_s$ and those to the numerator with two.  An inclusive calculation of $\gamma+j$ production,
such as the one we are considering in this paper, naturally contains terms in the numerator up to
$n_2 = n_1-1$.  Our NNLO calculation corresponds to $n_1 = 2$ while the equivalent result from our
NLO calculation is given by $n_1 = 1$.  We call these predictions
$R^{NNLO}_{2/1}$ and $R^{NLO}_{2/1}$ respectively and compare them to the CMS
measurement of the same ratio in Fig.~\ref{fig:njetratio}.
$R^{\rm{NLO}}_{2/1}(p_T^\gamma)$ does a poor job of describing the data because
it is a LO calculation for this observable and thus bears all the hallmarks of such a calculation.  This
is not only reflected by a general underestimation of the data, but also by the rather large scale dependence.
The corrections to this ratio when moving to $R^{\rm{NNLO}}_{2/1}(p_T^\gamma)$
are large, around 30\%. The agreement with data is significantly improved and the scale uncertainty is
reduced by a factor of two.

\begin{center}
\begin{figure}
\includegraphics[width=8cm]{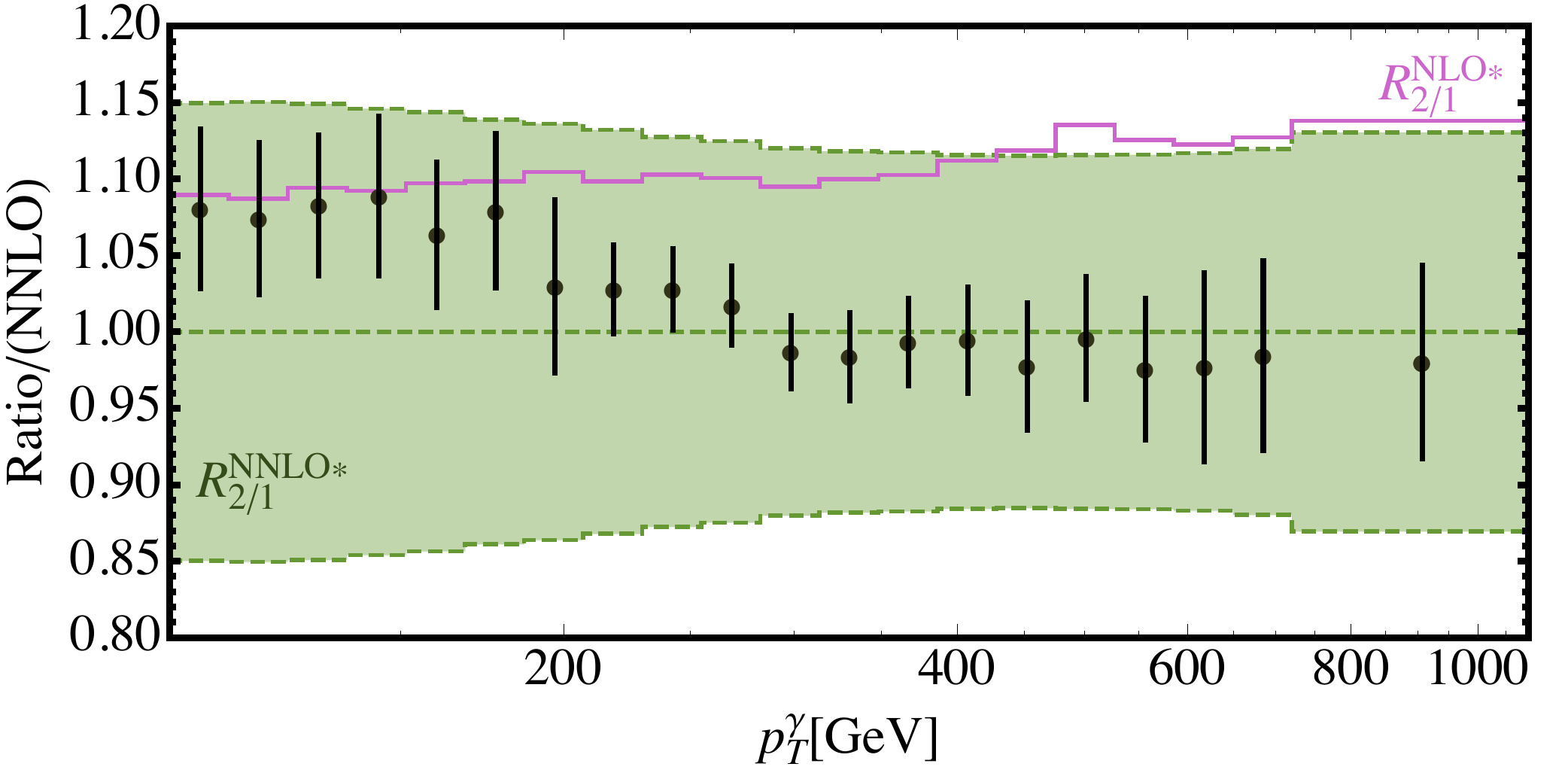}
\caption{The quantity $R_{2/1}^{\rm{NLO*}}(p_T^\gamma)$, a range of estimates for
$R_{2/1}^{\rm{NNLO*}}(p_T^\gamma)$ computed as discussed in the text, and the CMS
measurements~\cite{Khachatryan:2015ira}.  All quantities are normalized to the
theoretical prediction $R_{2/1}^{\rm{NNLO}}(p_T^\gamma)$.}
\label{fig:n2j1j_otherpred}
\end{figure}
\end{center}

However, from Eq.~(\ref{eq:R21defnew}) it is clear that neither of the predictions presented so far
corresponds to a strict expansion of the {\em ratio} to a given power of the
strong coupling, due to the fact that the denominator contains an additional term of one order higher than the numerator.
Instead we can define alternative predictions, corresponding to $n_2 = n_1$, with $R^{NLO\star}_{2/1}$
given by $n_1 = n_2 = 1$.   Note that the alternative definition $R^{NLO\star}_{2/1}$ can be obtained by
simply taking the ratio of two NLO calculations of $\gamma+2j$ and $\gamma+j$ production. This is
the procedure already followed by CMS~\cite{Khachatryan:2015ira} using the results of Ref.~\cite{Bern:2011pa}.  Since the NNLO corrections to $\gamma+2j$
production are unknown, and likely to remain so for some time, it is useful to estimate the potential
impact that they could have on the theoretical prediction for $R_{2/1}$.  We do so by postulating
NNLO corrections given by,
\begin{equation}
d\sigma_{\gamma+2j}^{(2,{\rm approx})}/dp_T^\gamma = \pm
 \frac{\left[d\sigma_{\gamma+2j}^{(1)}/dp_T^\gamma\right]^2}{d\sigma_{\gamma+2j}^{(0)}/dp_T^\gamma} \,,
\end{equation}
where, as indicated, the corrections can be of either sign.  In this way the NNLO corrections are of the
same size relative to NLO as the NLO ones are to LO.
A comparison of the results for $R^{NLO\star}_{2/1}$ and the two bounding estimates of $R^{NNLO\star}_{2/1}$,
with both our calculation of $R^{NNLO}_{2/1}$ and the CMS data, is shown in Fig.~\ref{fig:n2j1j_otherpred}.
We see that, as observed already in ref.~\cite{Khachatryan:2015ira}, the prediction
$R^{NLO\star}_{2/1}$ is in good agreement with the data for $p_T^\gamma < 200$~GeV but overshoots
it by around 15\% at high $p_T^\gamma$.  The range of the estimate $R^{NNLO\star}_{2/1}$ brackets both
the theory predictions $R^{NLO\star}_{2/1}$ and $R^{NNLO}_{2/1}$, as well as the data, and is of a
similar size as the scale uncertainty on $R^{NNLO}_{2/1}$ shown in Fig.~\ref{fig:njetratio}.  In addition,
the data suggests that NNLO corrections to $\gamma+2j$ production might be expected to be small at
high $p_T^\gamma$. In summary, $R^{NNLO}_{2/1}$ provides a fairly good description of the data
and we believe that the associated scale uncertainty provides a plausible envelope for the results of
a complete NNLO calculation of this ratio ($R^{NNLO*}_{2/1}$).

\section{The $Z/\gamma$ ratio at NNLO}
 
We are now able to address the principal aim of this paper, which is improving the theoretical prediction for
the ratio of $Z+j$ and $\gamma+j$ production.
We consider the case where the $Z$
boson decays to leptons and the two processes are studied in a similar kinematic regime by application
of the cuts described in section~\ref{sec:eventselection}.  Specifically, we consider predictions for
the quantity,
\begin{equation}
R^{\mathcal{O}}_{Z/\gamma}(p_T) =
 \frac{d\sigma^{\mathcal{O}}_{\ell^-\ell^++j+X}/dp_T}{d\sigma^{\mathcal{O}}_{\gamma+j+X}/dp_T} \,,
\label{eq:RZgdef}
\end{equation}
where $p_T$ represents the transverse momentum of the $Z$-boson or photon. 
A simple expectation for the behaviour of this ratio can be obtained by considering only the effect of the different
$Z$ and photon couplings, together with the effect of the PDFs, in the LO cross-section.  This neglects the effect of the
$Z$-boson mass, which should be irrelevant at large $p_T^Z$, as well as the impact of higher-order
corrections.  The ratio is then estimated to be~\cite{Ask:2011xf},
\begin{equation}
R_{Z/\gamma} = \left( R_u + \frac{R_d - R_u}{1 + \frac{Q_u^2}{Q_d^2}\frac{\langle u \rangle }{ \langle d \rangle}} \right)
 \left[  {\rm Br}(Z \to \ell^-\ell^+) \times {\cal A} \right] \,,
\label{eq:Rbehaviour}
\end{equation}
where $R_q$ is the relevant ratio of quark-boson couplings squared,
\begin{equation}
R_{q} = \frac{v_q^2 + a_q^2}{4 \sin^2\theta_w \cos^2 \theta_w Q_q^2} \,,
\end{equation}
and $\langle u \rangle$ ($\langle d \rangle$) is the typical up (down) quark PDF at the value of $x$ probed by
a given $p_T^V$, i.e. $\langle x \rangle = 2p_T^V / \sqrt{s}$.  The branching ratio and acceptance factor (${\cal A}$)
account for the $Z$-boson decay and cuts on the leptons.  At high transverse momentum, $p_T^V \gg M_Z$,
$x \to 1$ and  $\langle u \rangle / \langle d \rangle \to \infty$, so that $R_{Z/\gamma}$ should slowly approach
an asymptotic value from above~\cite{Bern:2011pa,Ask:2011xf}.   
This argument thus predicts a plateau at high transverse momentum, which we will observe shortly in our full prediction.
We stress that in our calculation this ratio is not computed for on-shell $Z$ bosons but includes the decay into
leptons, off-shell effects and the (small) contribution from virtual photon exchange.  Nevertheless, we will
refer to this quantity as $R_{Z/\gamma}$, or the $Z/\gamma$ ratio, as a matter of convenience.  


When computing this ratio a subtlety arises when trying to provide an uncertainty estimate based on
scale variation.  If the variation is correlated, i.e. one computes the scale uncertainty using
the same scale in both the numerator and denominator of Eq.~(\ref{eq:RZgdef}), then
one obtains essentially no uncertainty on $R_{Z/\gamma}(p_T)$, even at NLO.
We therefore discard this choice as a useful measure of the theoretical uncertainty.
The alternative that we use instead is to consider variations of the scale in
the numerator and denominator separately,
\begin{equation}
 \frac{d \sigma^{\mathcal{O}, \{r, f\}}_{\ell^-\ell^++j+X}/d p_T}
        {d \sigma^{\mathcal{O}, r=f=1}_{\gamma+j+X}/d p_T}
        \;\; \mbox{and} \;\; 
 \frac{d \sigma^{\mathcal{O}, r=f=1}_{\ell^-\ell^++j+X}/d p_T}
        {d \sigma^{\mathcal{O}, \{r, f\}}_{\gamma+j+X}/d p_T} \,,
\label{eq:RZgammavariation}
\end{equation} 
where $\{r,f\}$ represents the six-point scale variation indicated in Eq.~(\ref{eq:scalevariation}).
The uncertainty is then defined by the extremal values of either of these
two ratios.  In practice, since the scale-dependence of the two processes is so
similar, this procedure is almost identical to defining the uncertainty in
terms of the variation of either quantity in Eq.~(\ref{eq:RZgammavariation}) alone.
In contrast to the correlated variation, this approach results in scale uncertainties that,
order-by-order, overlap both the data and the central result of the next-higher order.
Moreover, with this procedure, at NNLO the resulting uncertainty band is of a size typical of
a NNLO prediction and still smaller than the experimental uncertainties. 

\begin{figure}
\includegraphics[width=8cm]{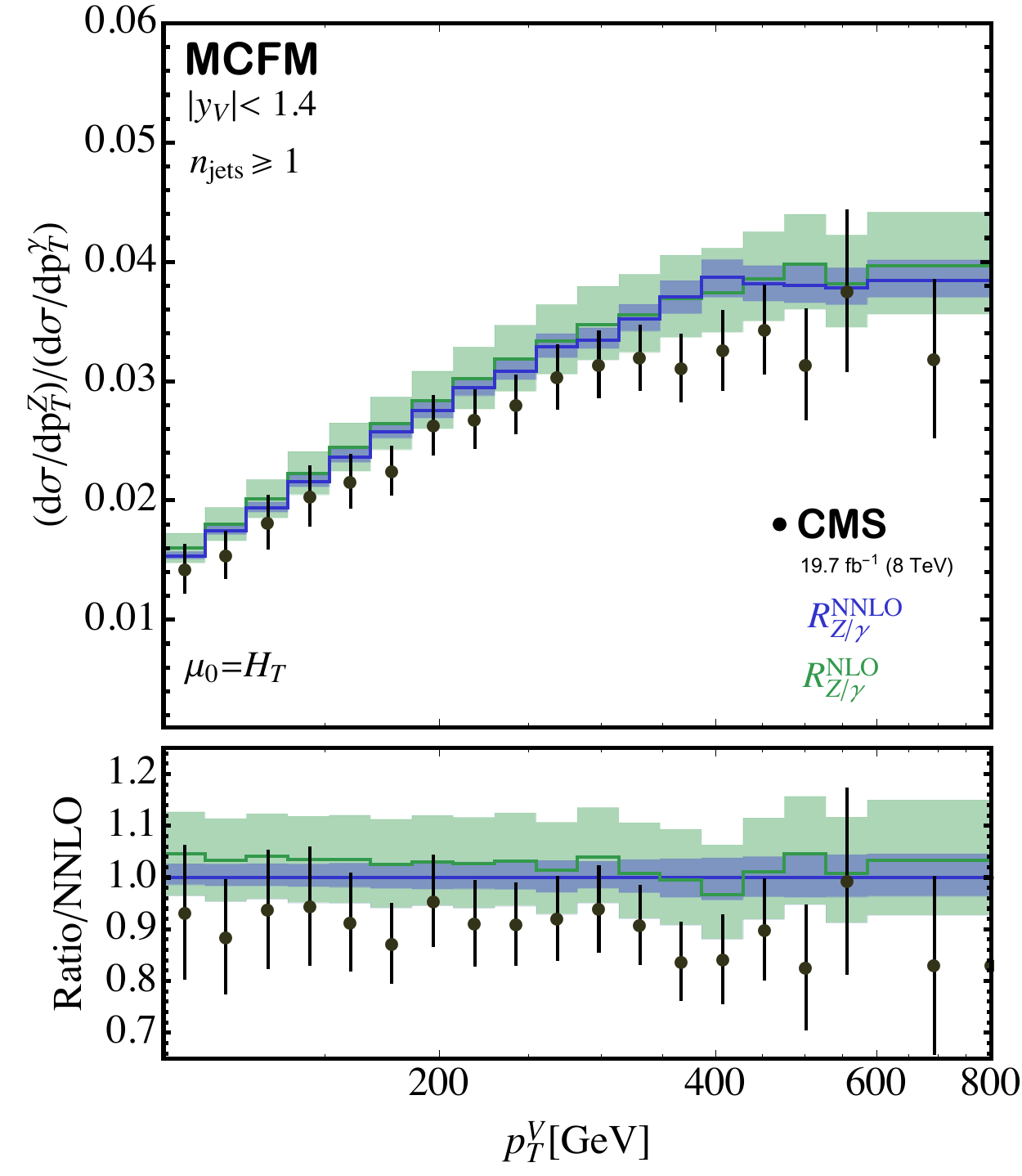}
\caption{The quantities $R_{Z/\gamma}^{\rm{NLO}}(p_T^\gamma)$ and $R_{Z/\gamma}^{\rm{NNLO}}(p_T^\gamma)$,
defined through Eq.~(\ref{eq:RZgdef}), compared to CMS data from ref.~\cite{Khachatryan:2015ira}.
The bands indicate the scale uncertainty on the theoretical predictions.}
\label{fig:ZgaRatio}
\end{figure}


\begin{figure}
\includegraphics[width=8cm]{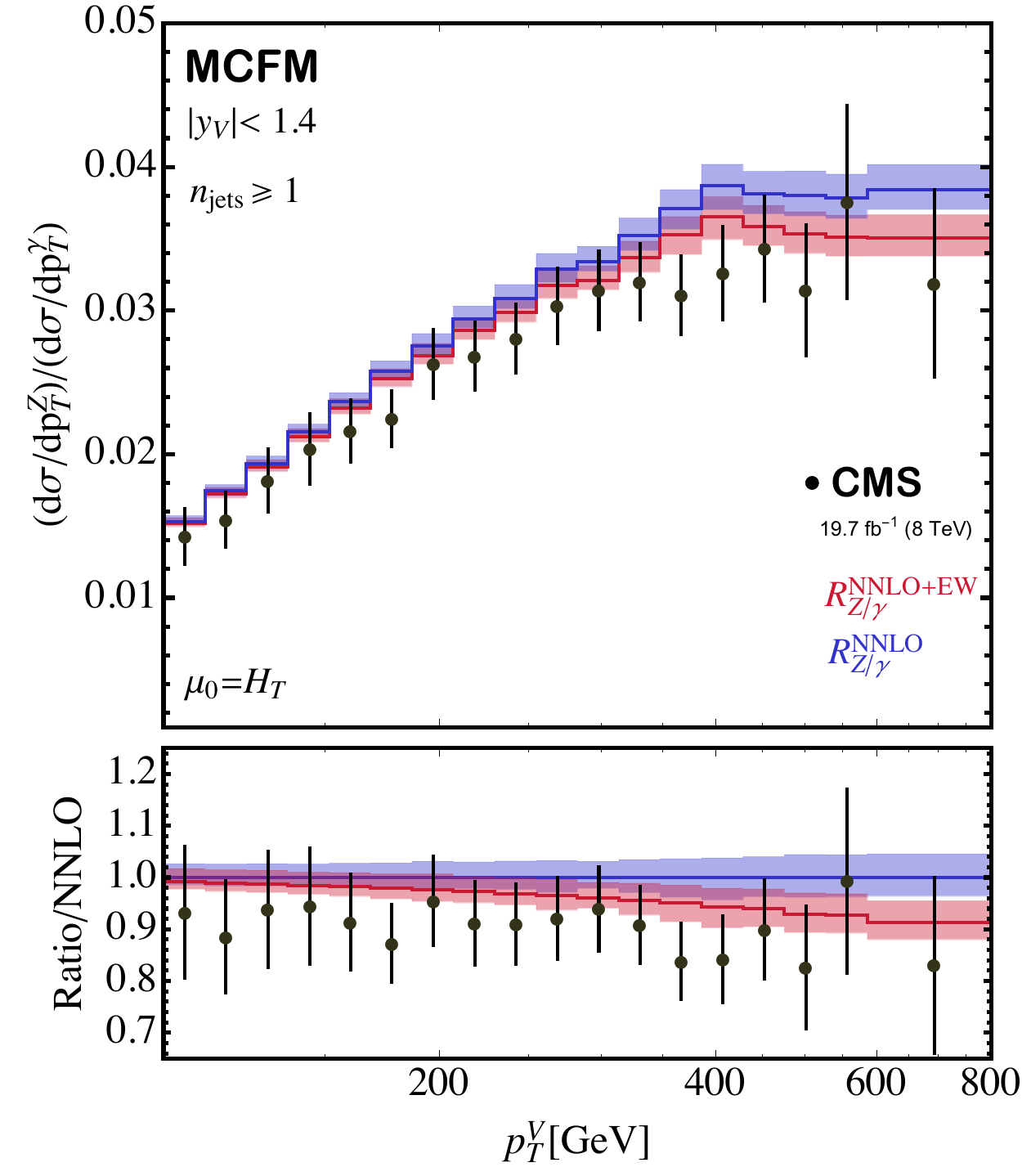}
\caption{The quantity $R_{Z/\gamma}(p_T^\gamma)$ defined in Eq.~(\ref{eq:RZgdef}), computed at NNLO
and at NNLO including EW effects, compared to CMS data from ref.~\cite{Khachatryan:2015ira}.  The bands
indicate the scale uncertainty on the theoretical predictions.}
\label{fig:ZgaRatioEW}
\end{figure}
 
Our results for the ratio for the pure QCD NLO and NNLO calculation are shown in 
Fig.~\ref{fig:ZgaRatio}.  The most significant effect of the NNLO calculation is to decrease the ratio,
particularly at lower values of $p_T$.
We have already seen, in Fig.~\ref{fig:ptgaNorm}, that the shape of the $p_T^\gamma$ spectrum is significantly improved
by the inclusion of electroweak effects.  We therefore extend our prediction for this ratio by taking such
corrections into account, rescaling the individual $p_T$ spectra by $(1+\Delta_{EW}^{V})$ as discussed previously.
Since the electroweak corrections do not affect the $Z+j$ and $\gamma+j$ processes in the same way~\cite{Kuhn:2005gv}, this leads to
a modification of the prediction for this ratio that is shown in Fig.~\ref{fig:ZgaRatioEW}.  Although the effects
are minor in the low-$p_T$ region, as expected, they become more important in the highest bins.  There they decrease
the ratio by as much as $7\%$ and thereby improve the agreement with the CMS data.

%

\begin{center}
\begin{figure}
\includegraphics[width=8cm]{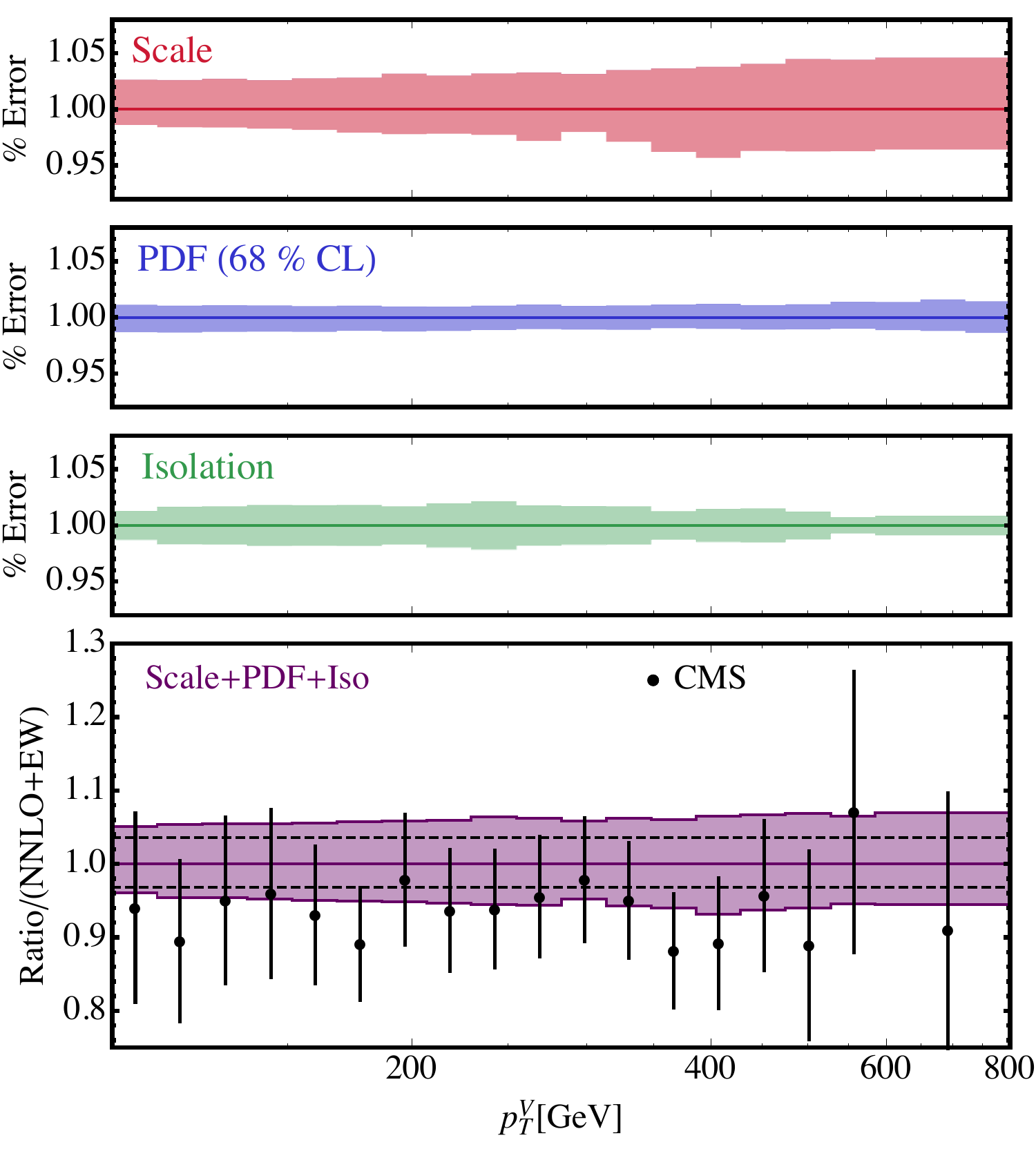}
\caption{A summary of the theoretical uncertainties discussed in this paper for the
$Z/\gamma$ ratio, $R^{NNLO+EW}_{Z/\gamma}$.  In order from the top, uncertainties from: scales, PDFs, isolation
and in the total, as described in the main text. The total uncertainty
is obtained by combining linearly those from the sources above.  The uncertainty due to
the value of $\alpha_{EM}$ is indicated separately by the dashed line in the lower figure.}
\label{fig:pdfuncert_Zga}
\end{figure}
\end{center}

We now consider a full analysis of the theoretical uncertainties associated with the
calculation of $R^{NNLO+EW}_{Z/\gamma}$.  We use the same procedure as discussed earlier for
the photon $p_T$ spectrum, except that we only vary $\alpha_{EM}$ in the $\gamma+j$ prediction
and fix $\alpha = 1/132.232$ in the $Z+j$ calculation.
Our results are presented in Figure~\ref{fig:pdfuncert_Zga} where, as before,
the uncertainties are normalized to the central value of the combined NNLO QCD + EW prediction.  
We see that the PDF uncertainties essentially cancel, as one might expect from
the nature of the ratio. The dominant uncertainties are those resulting
from scale variation (especially at high $p_T^V$) and a change in the overall normalization from $\alpha_{EM}$.
The total uncertainty is only around 4\% in the lowest bins and is slightly higher,
approximately 6\%, at high $p_T^\gamma$.

As discussed earlier, the asymptotic behavior of our prediction for $R_{Z/\gamma}$ is particularly
interesting.  In order to quantify this we follow the CMS analysis~\cite{Khachatryan:2015ira} and
define a ratio in which the high-$p_T$ bins are integrated over,
\begin{eqnarray}
R_{{\rm{dilep}}} = \frac{\sigma_{\ell^-\ell^+ + j + X}(p_T^V > 314 \; {\rm{GeV}})}{\sigma_{\gamma+j+X}(p_T^V > 314 \; {\rm{GeV}})} \,.
\label{eq:Rdilepdefn}
\end{eqnarray}
The experimental measurement of this quantity by CMS is,
\begin{eqnarray}
R^{{\rm CMS}}_{{\rm{dilep}}}  &=& 0.0322 \pm 0.0008 \; (\rm{stat}) \pm 0.0020\; (\rm{syst}) \,. \nonumber
\end{eqnarray}
Our best theoretical prediction is provided by the NNLO+EW prediction shown in Figure~\ref{fig:ZgaRatioEW},
with accompanying uncertainties illustrated in Figure~\ref{fig:pdfuncert_Zga}.  We find,
\begin{eqnarray}
&& R^{{\rm NNLO+EW}}_{{\rm{dilep}}}(8 \;\rm{TeV}) =  0.0348 \nonumber \\
&&
{}_{-0.0013}^{+0.0012} \; (\rm{scale})\, {}_{-0.0004}^{+0.0004} \; (\rm{PDF}) \, 
{}_{-0.0006}^{+0.0006} \; (\rm{iso}) \, {}_{-0.0012}^{+0.0012} \; (\alpha_{EM}) \,. \nonumber
\end{eqnarray}
This result is in excellent agreement with the measured value, $R^{{\rm CMS}}_{{\rm{dilep}}}$.



The CMS collaboration has not yet performed a similar analysis of $\gamma+j$ production
at 13 TeV.  Since such an undertaking will likely involve a change in the cuts that are
applied, or at least in the binning of the final data, for now we refrain from performing a
detailed study of individual distributions at this energy.  However it is especially important
to predict the ratio $R_{Z/\gamma}(p_T)$ and, in particular, its value in the high-$p_T$ tail.
For this reason we repeat our above analysis at 13 TeV, with no cuts or input parameters altered
apart from the LHC operating energy.

\begin{center}
\begin{figure}
\includegraphics[width=8cm]{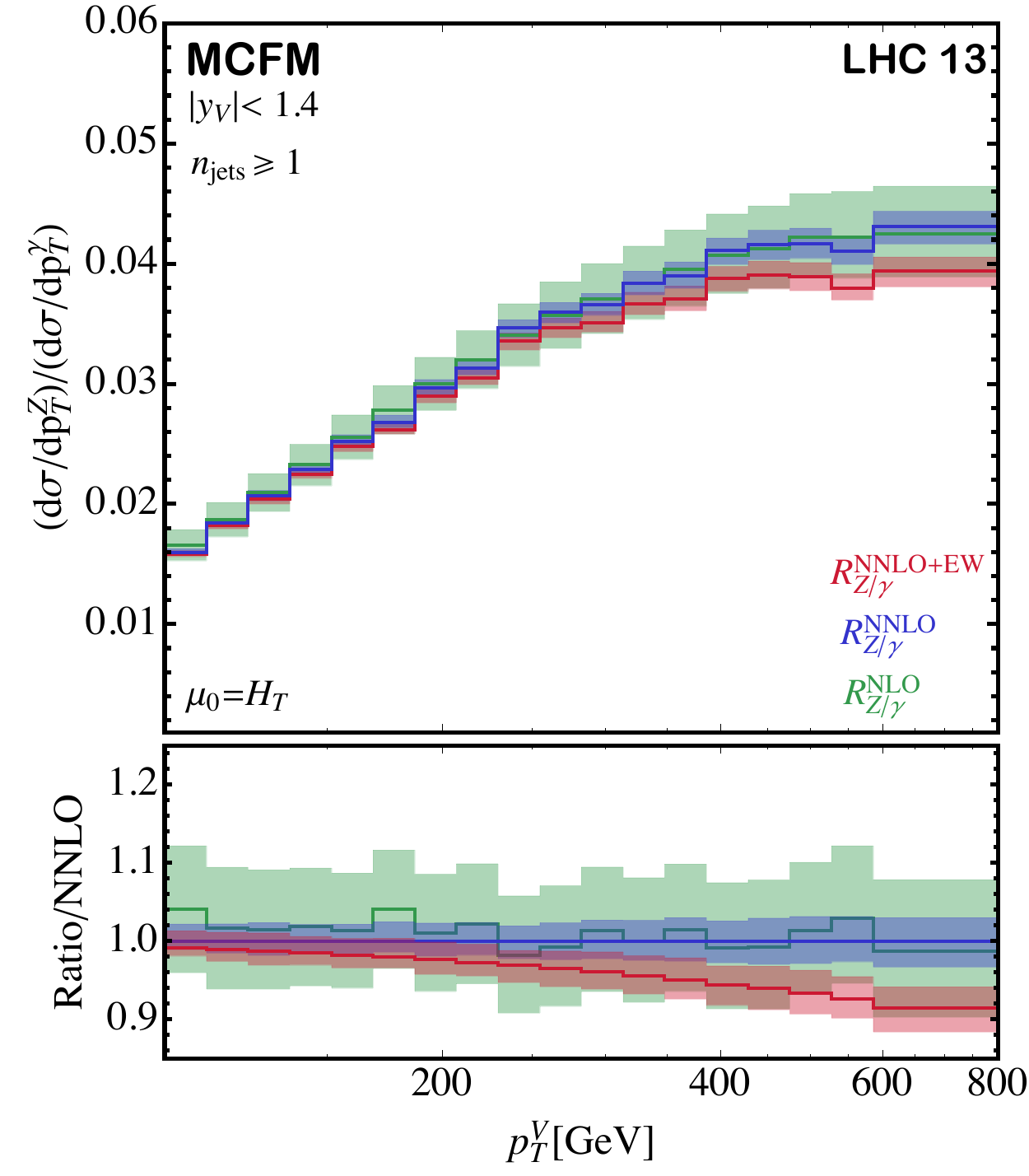}
\caption{The quantities $R_{Z/\gamma}^{\rm{NLO}}(p_T^\gamma)$, $R_{Z/\gamma}^{\rm{NNLO}}(p_T^\gamma)$
and $R_{Z/\gamma}^{\rm{NNLO+EW}}(p_T^\gamma)$, defined through Eq.~(\ref{eq:RZgdef}), for the LHC
operating at 13 TeV. The bands indicate the scale uncertainty on the theoretical predictions.}
\label{fig:pdfuncert13}
\end{figure}
\end{center}

Our prediction for $R_{Z/\gamma}(p_T)$ at 13 TeV is shown in Figure~\ref{fig:pdfuncert13}, where
we compare predictions at NLO, NNLO and when combining NNLO QCD and EW effects. As before
(c.f. Figures~\ref{fig:ZgaRatio} and~\ref{fig:ZgaRatioEW}) we see that the ratio is very similar
in all cases, but that the NNLO prediction has a substantially smaller uncertainty and the
inclusion of EW effects lowers the ratio at high $p_T$.
At 13 TeV we are further from the large-$x$ region, for the same range of $p_T^\gamma$,
so that the $\langle u \rangle / \langle d \rangle$ ratio in Eq.~(\ref{eq:Rbehaviour}) is smaller.
We thus expect that the value of $R_{\rm{dilep}}$ is higher at $13$~TeV than at $8$~TeV, a supposition
that is borne out by our explicit calculations.  We find, for the asymptotic ratio defined in
Eq.~(\ref{eq:Rdilepdefn}),
\begin{eqnarray}
&& R^{{\rm NNLO+EW}}_{{\rm{dilep}}}(13 \;\rm{TeV}) =  0.0377 \nonumber \\
&&
{}_{-0.0011}^{+0.0013} \; (\rm{scale}) \, {}_{-0.0004}^{+0.0004} \; (\rm{PDF})  \,
{}_{-0.0006}^{+0.0006} \; (\rm{iso}) \, {}_{-0.0013}^{+0.0013} \; (\alpha_{EM}) \,. \nonumber
\end{eqnarray}

\begin{center}
\begin{figure}
\includegraphics[width=8cm]{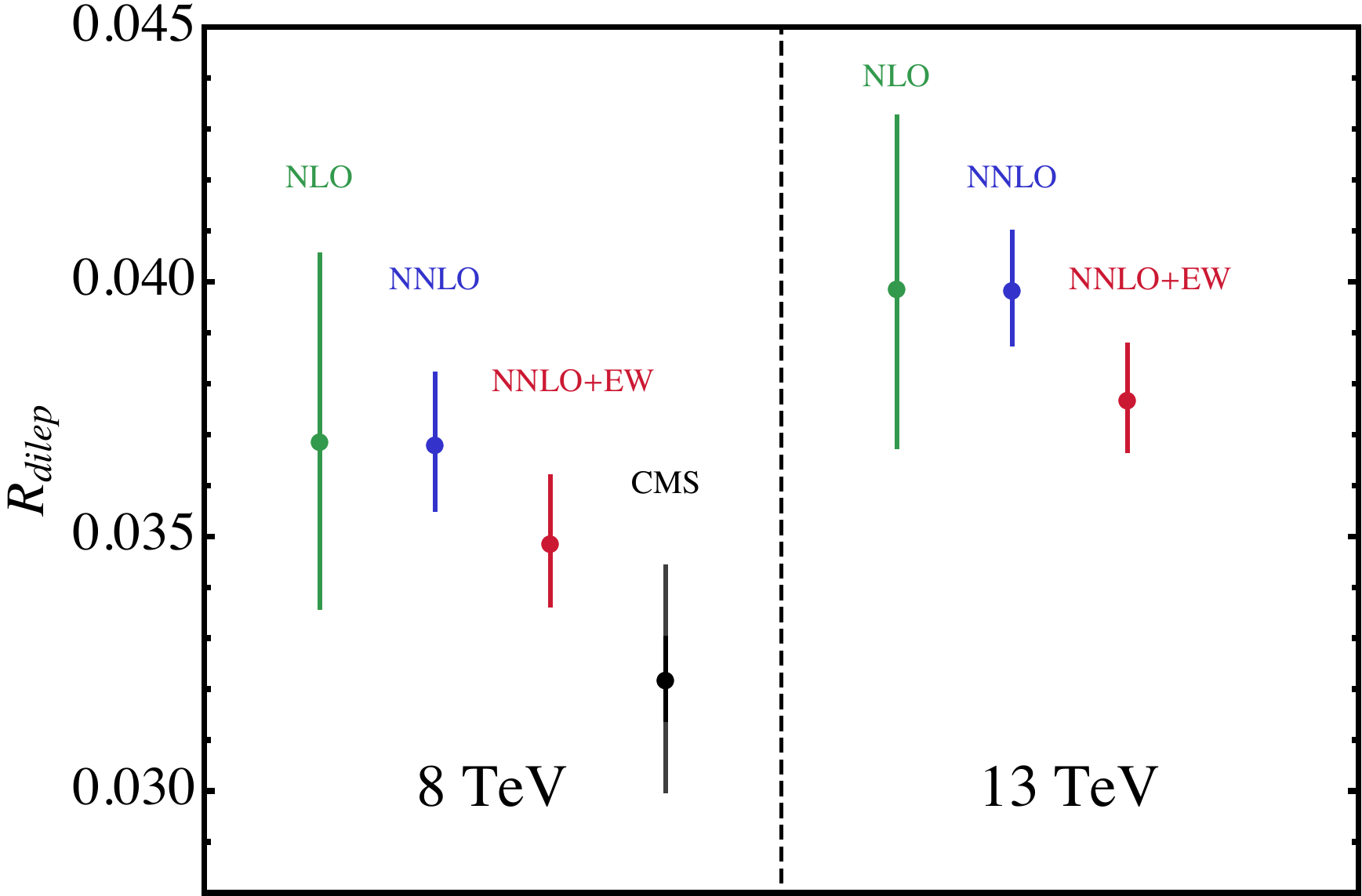}
\caption{A summary of predictions for, and measurements of, $R_{{\rm{dilep}}}$ -- defined in Eq.~(\ref{eq:Rdilepdefn})
 -- at 8 and 13 TeV.}
\label{fig:Rdilep}
\end{figure}
\end{center}

We conclude this section with a summary of the theoretical predictions for $R_{\rm dilep}$, computed
at various orders of perturbation theory, shown in Figure~\ref{fig:Rdilep}. 
The improvement in the precision of the theoretical prediction  when going from NLO to NNLO QCD is clear.
It also emphasizes that, after the inclusion of electroweak effects, there is excellent agreement
between the best theoretical prediction and the measurement of CMS~\cite{Khachatryan:2015ira}.

\section{Conclusions}
\label{sec:conc}

In this paper we have presented differential predictions for $\gamma+j$ production at NNLO
and compared our predictions to data taken by the CMS experiment at 8 TeV.  We have
seen that NNLO predictions provide a very good description of the shape of the CMS data,
with the inclusion of EW effects improving the agreement further still. For the
$p_T^\gamma$ distribution there is an apparent disagreement between the normalization
of the theoretical prediction and the observed data, but again the shapes of the theory
and data are very similar.  We have used our results to compute several other quantities, notably the ratio of
$Z+j$ and $\gamma+j$ production as a function of the boson transverse momentum, which is useful for
estimating backgrounds to BSM searches.  The agreement between the theoretical prediction and data
for this ratio is excellent.  Finally, we have made additional predictions at NNLO accuracy for future studies of the
$Z+j$/$\gamma+j$ ratio at 13 TeV.

\section*{Acknowledgements}

We thank Joe Incandela, Jonas Lindert, Michelangelo Mangano and Ruth Van de Water for useful discussions. 
Support provided by the Center for Computational Research at the University at Buffalo
and the Wilson HPC Computing Facility at Fermilab.
CW is supported by the National Science Foundation through award number PHY-1619877.
The research of JMC is supported by the US DOE under contract DE-AC02-07CH11359.


\section*{Appendix}

Our numerical results for the $Z+j$ to $\gamma+j$ ratio studied in this paper, $R_{Z/\gamma}$, are
presented in Table~\ref{tab:ptVrat} (8 TeV) and Table~\ref{tab:ptVrat13} (13 TeV). For each bin of
$p_T^V$ we show the value of the ratio computed to NLO and NNLO accuracy, the associated uncertainty due
to scale variation as described in the text, and the EW rescaling factor.  

\begin{table}
\begin{tabular}{|c|c|c|c|}
\hline
$p_T^{V}$ [GeV] \; & $R_{Z/\gamma}^{\rm NLO} \times 100 \; $ & $R_{Z/\gamma}^{\rm NNLO} \times 100 \; $ &
 $\frac{1+\Delta^Z_{\rm{EW}}}{1+\Delta^\gamma_{\rm{EW}}}$ \\
\hline
\hline
$ \text{100-111} $ & $1.60_{-0.20}^{+0.23} $ & $1.53_{-0.04}^{+0.05} $ & $0.99  $ \\
$ \text{111-123} $ & $1.80_{-0.23}^{+0.26} $ & $1.74_{-0.05}^{+0.05} $ & $0.99  $ \\
$ \text{123-137} $ & $2.01_{-0.27}^{+0.3} $ & $1.94_{-0.06}^{+0.06} $ & $0.99  $ \\
$ \text{137-152} $ & $2.23_{-0.31}^{+0.34} $ & $2.15_{-0.07}^{+0.07} $ & $0.98  $ \\
$ \text{152-168} $ & $2.45_{-0.34}^{+0.39} $ & $2.36_{-0.08}^{+0.09} $ & $0.98  $ \\
$ \text{168-187} $ & $2.64_{-0.38}^{+0.43} $ & $2.58_{-0.10}^{+0.10} $ & $0.98  $ \\
$ \text{187-207} $ & $2.84_{-0.41}^{+0.48} $ & $2.75_{-0.11}^{+0.12} $ & $0.98  $ \\
$ \text{207-230} $ & $3.02_{-0.45}^{+0.52} $ & $2.94_{-0.11}^{+0.12} $ & $0.97  $ \\
$ \text{230-255} $ & $3.18_{-0.48}^{+0.56} $ & $3.08_{-0.12}^{+0.13} $ & $0.97  $ \\
$ \text{255-283} $ & $3.34_{-0.52}^{+0.6} $ & $3.29_{-0.15}^{+0.16} $ & $0.96  $ \\
$ \text{283-314} $ & $3.47_{-0.55}^{+0.64} $ & $3.34_{-0.12}^{+0.12} $ & $0.96  $ \\
$ \text{314-348} $ & $3.55_{-0.57}^{+0.68} $ & $3.52_{-0.16}^{+0.18} $ & $0.96  $ \\
$ \text{348-386} $ & $3.69_{-0.61}^{+0.73} $ & $3.71_{-0.21}^{+0.24} $ & $0.95  $ \\
$ \text{386-429} $ & $3.74_{-0.63}^{+0.75} $ & $3.87_{-0.25}^{+0.28} $ & $0.94  $ \\
$ \text{429-476} $ & $3.86_{-0.67}^{+0.8} $ & $3.81_{-0.23}^{+0.24} $ & $0.94  $ \\
$ \text{476-528} $ & $3.98_{-0.71}^{+0.86} $ & $3.80_{-0.25}^{+0.26} $ & $0.93  $ \\
$ \text{528-586} $ & $3.81_{-0.68}^{+0.83} $ & $3.78_{-0.24}^{+0.25} $ & $0.93  $ \\
$ \text{586-800} $ & $3.97_{-0.75}^{+0.93} $ & $3.84_{-0.24}^{+0.23} $ & $0.91  $ \\
 \hline
\end{tabular}
\caption{The values of $R_{Z/\gamma}^{\rm NLO}$ and $R_{Z/\gamma}^{\rm NNLO}$ at 8 TeV  (rescaled by a factor of 100),
together with the additional correction that corresponds to including EW effects in both processes. Quoted ranges correspond to the variation in the central scale by the six-point method described in the text. 
}
\label{tab:ptVrat}
\end{table}

\begin{table}
\begin{tabular}{|c|c|c|c|}
\hline
$p_T^{V}$ [GeV] \; & $R_{Z/\gamma}^{\rm NLO} \times 100 \; $ & $R_{Z/\gamma}^{\rm NNLO} \times 100 \; $ &
 $\frac{1+\Delta^Z_{\rm{EW}}}{1+\Delta^\gamma_{\rm{EW}}}$ \\
\hline
\hline
$ \text{100-111} $ & $1.66_{-0.21}^{+0.23} $ & $1.59_{-0.03}^{+0.04} $ & $0.99 $ \\
$ \text{111-123} $ & $1.87_{-0.24}^{+0.26} $ & $1.84_{-0.04}^{+0.05} $ & $0.99 $ \\
$ \text{123-137} $ & $2.10_{-0.26}^{+0.29} $ & $2.07_{-0.06}^{+0.07} $ & $0.99 $ \\
$ \text{137-152} $ & $2.33_{-0.29}^{+0.32} $ & $2.28_{-0.06}^{+0.06} $ & $0.98 $ \\
$ \text{152-168} $ & $2.56_{-0.32}^{+0.35} $ & $2.52_{-0.06}^{+0.07} $ & $0.98 $ \\
$ \text{168-187} $ & $2.78_{-0.35}^{+0.39} $ & $2.67_{-0.07}^{+0.08} $ & $0.98 $ \\
$ \text{187-207} $ & $3.00_{-0.38}^{+0.43} $ & $2.97_{-0.09}^{+0.10} $ & $0.98 $ \\
$ \text{207-230} $ & $3.20_{-0.42}^{+0.47} $ & $3.13_{-0.10}^{+0.10} $ & $0.97 $ \\
$ \text{230-255} $ & $3.40_{-0.46}^{+0.52} $ & $3.47_{-0.10}^{+0.12} $ & $0.97 $ \\
$ \text{255-283} $ & $3.57_{-0.49}^{+0.56} $ & $3.59_{-0.13}^{+0.14} $ & $0.96 $ \\
$ \text{283-314} $ & $3.70_{-0.52}^{+0.59} $ & $3.66_{-0.14}^{+0.14} $ & $0.96 $ \\
$ \text{314-348} $ & $3.83_{-0.54}^{+0.63} $ & $3.84_{-0.15}^{+0.16} $ & $0.96 $ \\
$ \text{348-386} $ & $3.95_{-0.57}^{+0.66} $ & $3.90_{-0.17}^{+0.17} $ & $0.95 $ \\
$ \text{386-429} $ & $4.07_{-0.6}^{+0.7} $ & $4.11_{-0.16}^{+0.17} $ & $0.94 $ \\
$ \text{429-476} $ & $4.12_{-0.62}^{+0.73} $ & $4.16_{-0.19}^{+0.20} $ & $0.94 $ \\
$ \text{476-528} $ & $4.22_{-0.64}^{+0.75} $ & $4.17_{-0.19}^{+0.20} $ & $0.93 $ \\
$ \text{528-586} $ & $4.22_{-0.66}^{+0.78} $ & $4.10_{-0.21}^{+0.18} $ & $0.93 $ \\
$ \text{586-800} $ & $4.25_{-0.69}^{+0.81} $ & $4.31_{-0.24}^{+0.22} $ & $0.91 $ \\
 \hline
\end{tabular}
\caption{The values of $R_{Z/\gamma}^{\rm NLO}$ and $R_{Z/\gamma}^{\rm NNLO}$ at 13 TeV (rescaled by a factor of 100),
together with the additional correction that corresponds to including EW effects in both processes. Quoted ranges correspond to the variation in the central scale by the six-point method described in the text. 
}
\label{tab:ptVrat13}
\end{table}


\bibliography{gamjetCMS} 

\end{document}